\documentclass[aps,nofootinbib,prl,superscriptaddress,tightenlines,notitlepage,twocolumn,showpacs,floatfix]{revtex4-1}
\usepackage[colorlinks]{hyperref}
\usepackage{tabularx}
\usepackage{graphicx}
\usepackage{amssymb,bm,tensor}
\usepackage{textcomp}
\usepackage{xcolor}
\usepackage{amsmath}
\usepackage[varg]{txfonts}
\usepackage{enumerate}
\usepackage{mathtools}
\usepackage[normalem]{ulem}
\usepackage{bm} 
\usepackage[OT1]{fontenc}
\usepackage{aas_macros}
\usepackage{pdfpages}
\usepackage{etoolbox} 
\makeatletter
\patchcmd{\@outputpage@head}{\@ifx{\LS@rot\@undefined}{}{\LS@rot}}{}{}{}
\makeatother

\newcommand{\AIfA}{\affiliation{Argelander Institut f\"ur Astronomie, Auf
dem H\"ugel 71, D-53121 Bonn, Germany}}
\newcommand{\MPIfR}{\affiliation{Max-Planck-Institut f\"ur Radioastronomie, Auf
dem H\"ugel 69, D-53121 Bonn, Germany}}
\newcommand{\Aarhus}{\affiliation{Department of Physics and Astronomy, Aarhus University, Ny Munkegade 120, 8000 Aarhus C, Denmark}}

\begin{document}

\title{Disentangling Coalescing Neutron Star--White Dwarf Binaries for \textit {LISA}}
\date{\today}
\author{Thomas M. Tauris}\email{tauris@phys.au.dk}\AIfA\MPIfR\Aarhus

\begin{abstract}
The prime candidate sources for the upcoming space-borne gravitational wave (GW) observatory \textit{LISA} are the numerous Galactic tight binaries of white dwarfs (WDs) and neutron stars (NSs), many of which will coalesce and undergo mass transfer, leading to simultaneous emission of X-rays and GWs. Here, detailed and coherent numerical stellar models are explored for the formation and evolution of these systems, including finite-temperature effects and complete calculations of mass transfer from a WD to a NS accretor. Evolutionary tracks of characteristic strain amplitude are computed, and the unique pattern of their evolution in the GW frequency--dynamical chirp mass parameter space enables a firm identification of the nature of the systems. Furthermore, it is demonstrated that a precise detection of the chirp allows determination of the NS mass to an accuracy of a few per~cent, with applications to constraining its equation-of-state, in particular for dual-line GW sources observed simultaneously at high and low frequencies.
\end{abstract}

\maketitle


\paragraph{Introduction.}
The recent detections of high-frequency gravitational waves (GWs) from mergers of black holes (BHs) and neutron stars (NSs) in distant galaxies~\cite{aaa+16,aaa+17c} have excited the scientific community and marks the start of a new era of multi-messenger astrophysics. Sources of continuous emission of low-frequency GWs, however, are numerous within the Milky Way~\cite{nyp01}. These sources include mainly tight binaries of compact objects: white dwarfs (WDs), NSs and BHs. As these compact objects orbit each other and produce ripples in the local space time, GWs are emitted which result in a gradual orbital decay over time.
This causes a chirp of the emitted GW signal, which is an increase in frequency and amplitude, reaching a maximum when the two compact objects finally merge. A space-borne GW observatory (LISA~\cite{aab+17}) is planned for launch in about a decade, with an aim of detecting the chirp signals from such low-frequency GW sources. This opens up for the possibility to explore full multi-messenger detections in both GWs and electromagnetic waves from such tight binaries in which stable mass transfer (leading to emission of X-rays) is operating between the two compact objects, e.g. from a low-mass helium WD donor to a NS or WD accretor.
More massive carbon-oxygen WD donors are not considered here as their mass transfer is dynamically unstable~\cite{vnv+12}. 

Vigorous studies are known in the literature on WD\,+\,WD evolution~\citep[e.g.][]{hb00,nyp01,kbs12,kblk17}.
However, thus far attempts to model the chirp of the emitted GW signal are based on semi-analytic modelling, with limited possibilities to resolve finite-temperature (entropy) effects of the WD and the stability of the mass-transfer process. Here, the aim is to expand beyond semi-analytical results by using numerical modelling and investigate GW calculations of NS\,+\,WD systems for the first time.
An advantage of applying state-of-the-art numerical calculations is that one is not restricted to applications of approximate zero-temperature mass-radius relations of the WD, which therefore results in more realistic UCXB modelling~\cite{stli17}. This is particularly important for the low-mass helium WD donors studied here, since they can remain bloated on a Gyr timescale~\cite{itla14} until they settle on the WD cooling track. 
Finally, the ability to follow the coherent evolution of the same system through two consecutive mass-transfer stages leads to a self-consistent modelling of the WD donor. 

\paragraph{Binary star modelling.}
Using the numerical binary stellar evolution tool MESA~\cite{pms+15}, the complete evolution of NS binaries with a low-mass main sequence (MS) companion star is calculated until a double compact object is formed, and beyond (see Supplemental Material~\cite{SM} for further details on the calculations). This includes two consecutive stages of mass transfer: (i) the low-mass X-ray binary (LMXB) stage~\cite{lvv07} where the NS accretes matter from the MS donor star, and (ii) the ultra-compact X-ray binary (UCXB) stage~\cite{vnv+12,hie+13} where the NS accretes matter from the WD remnant of the former MS star. 
The computation of the UCXB stage, which had not been calculated numerically until recently~\cite{stli17}, holds the key for tracking the observable properties of such systems in both GWs and electromagnetic waves.

The example shown in Figure~1 is based on an initial binary with a 1.40\,M\,$_\odot$ MS star orbiting a 1.30\,M\,$_\odot$ NS with an orbital period of $P\simeq$~3.0\,days. After orbital decay caused by magnetic braking and the subsequent LMXB phase, the system detaches with a 0.162\,M\,$_\odot$ helium WD orbiting a 1.63\,M\,$_\odot$ NS with an orbital period of $4.8\;{\rm hr}$. 
At this stage, the system is observable as a binary radio millisecond pulsar (MSP, a recycled NS~\cite{bv91}). Over the next $\sim 3$~Gyr, the system spirals~in further due to emission of low-frequency gravitational waves (GWs) with a constant chirp mass, $\mathcal{M}=(M_{\rm NS}M_{\rm WD})^{3/5}/(M_{\rm NS}+M_{\rm WD})^{1/5}=0.401$\,M\,$_\odot$, until the WD fills its Roche lobe (at $P$\,=\,24~min and with a temperature of $T_{\rm eff}$\,=\,10\,580~K) and initiates mass transfer (Roche-lobe overflow, RLO) to the NS and the system becomes observable as an UCXB.
It is anticipated that a large subpopulation of LISA sources~\cite{nyp01,aab+17} will indeed be such X-ray binaries, where mass is transferred from a low-mass WD to a NS. The sample system shown here is calculated until an age of $\sim$14~Gyr at which point the WD has become a $\sim\!0.006$\,M\,$_\odot$ planet-like remnant orbiting an MSP -- somewhat similar to a system like PSR~J1719$-$1438 \citep{bbb+11}. 
\begin{figure}
  \vspace{-0.7cm}\includegraphics[width=10.5cm]{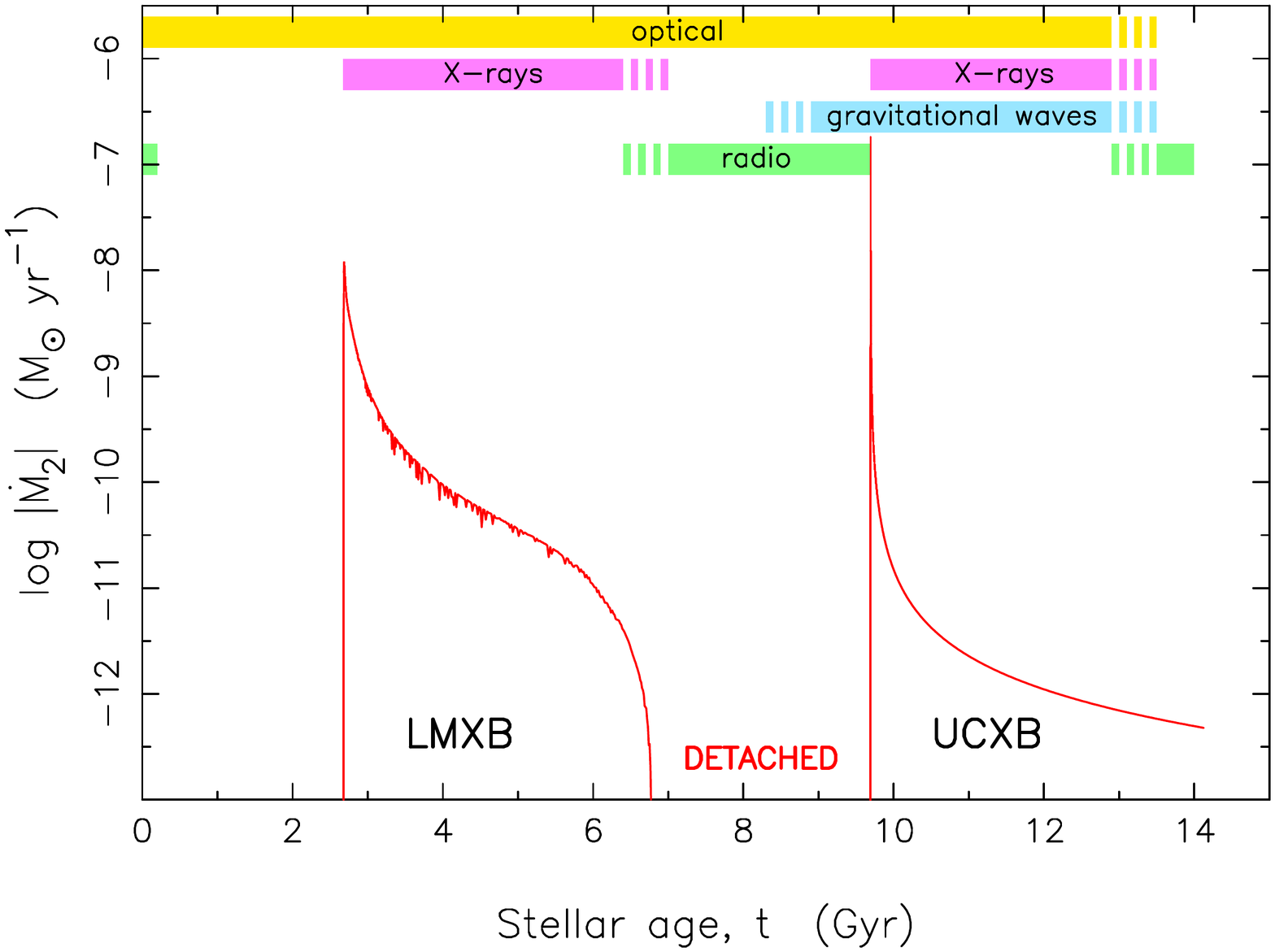}
  \caption{\label{fig1} Mass-transfer rate of the donor star as a function of stellar age. The initial MS star\,+\,NS binary has components of 1.40\,M$_\odot$ and 1.30\,M$_\odot$, respectively. The system evolves through two observable stages of mass transfer:
an LMXB for 4\,Gyr, followed by a detached phase lasting about 3\,Gyr where the system is detectable as a radio millisecond pulsar orbiting the helium WD remnant of the donor star, until GW radiation brings the system into contact again, producing an UCXB. The colour bars indicate detectability in different regimes.}
\end{figure}
\begin{figure*}
  \vspace{-0.7cm}\includegraphics[width=10cm]{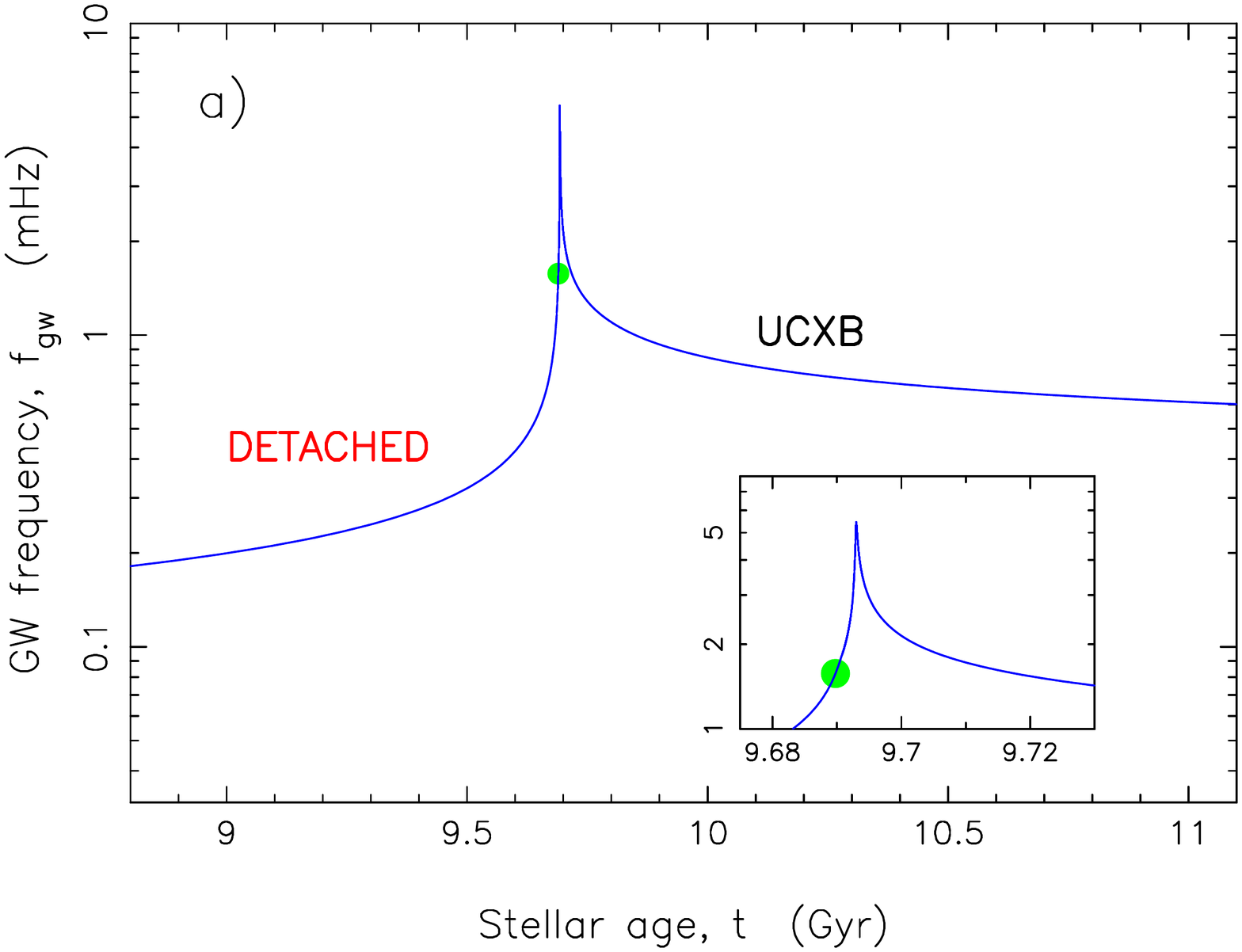}\includegraphics[width=10cm]{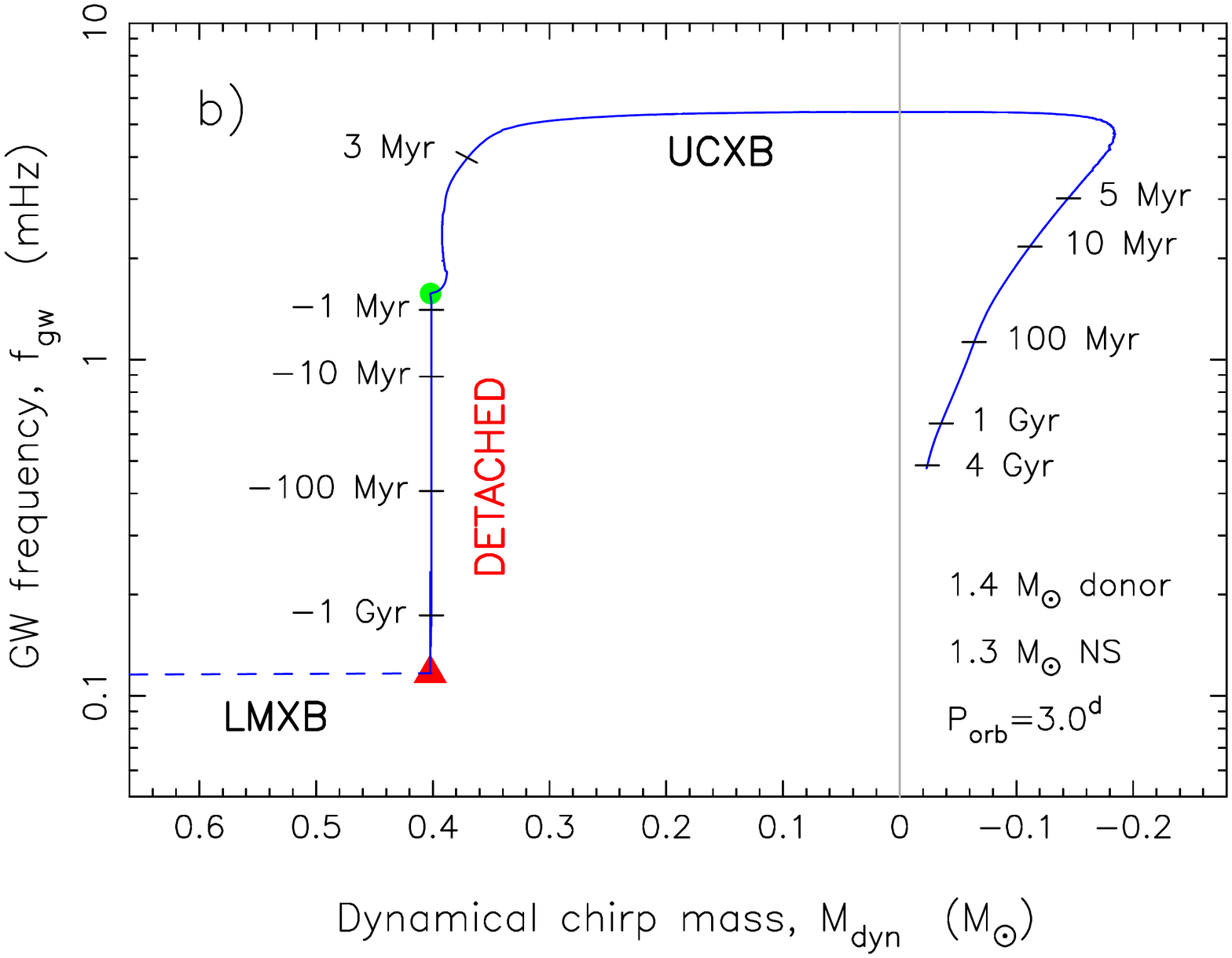} 
  \caption{\label{fig2} Calculated GW spectrum evolution during the mass transfer from a 0.162\,M$_\odot$ WD to a 1.63\,M$_\odot$ NS in the UCXB of Figure~1. {\bf a)}~Emitted GW frequency vs stellar age. The insert shows a zoom-in near the peak frequency. {\bf b)}~Emitted GW frequency vs dynamical chirp mass. The evolution is from left to right in a clock-wise direction. The ages along the evolutionary track are relative to the onset of the UCXB phase (green solid circle) at t\,=\,9.690~Gyr. The dynamical chirp mass becomes negative when the orbit is widening as a result of mass transfer.}
\end{figure*}

\paragraph{Dynamical chirp mass.}
Figure~2 displays the calculated GW frequency, $f_{\rm gw}$ as a function of stellar age (left panel) and so-called \textit{dynamical} chirp mass, $\mathcal{M}_{\rm dyn}$ (right panel), before and after the onset of the UCXB stage. These tracks represent a unique fingerprint of GW frequency evolution (or a ``song'') for a given binary system.
In the quadrupolar formalism, $f_{\rm gw}$ is simply twice the orbital frequency ($f=1/P$) and
the latter quantity, $\mathcal{M}_{\rm dyn}$ depends on $f$ and its time derivative, $\dot{f}$. 

For a detached binary system (i.e. without mass transfer between the two stellar components) where the only contribution to loss of orbital angular momentum is caused by GW radiation, the chirp mass is a constant quantity defined as~\cite{mag08}:
\begin{equation}
\mathcal{M} \equiv \frac{(M_1\,M_2)^{3/5}}{(M_1+M_2)^{1/5}}
\end{equation}
Given that the loss rate of orbital angular momentum caused by GW radiation for a circular binary can be expressed as~\cite{pet64}:
\begin{equation}
  \frac{\dot{J}_{\rm gwr}}{J_{\rm orb}} = - \frac{32\,G^3}{5\,c^5}\,\frac{M_1 M_2 M}{a^4}
\label{eq:Jdotgw}
\end{equation}
where $J_{\rm orb}$ is the orbital angular momentum, $G$ is the constant of gravity, $c$ is the speed of light in vacuum, $M=M_1+M_2$ is the total mass of the system, and $a$ is the orbital separation between the stellar components, one can combine the above expression with Kepler's third law ($4\pi^2 f^2 = GM/a^3$, where $f=f_{\rm gw}/2\,$ is the orbital frequency), and easily derive:
\begin{equation}
\mathcal{M} = \frac{c^3}{G}\,\left(\frac{5}{96}\pi^{-8/3}f_{\rm gw}^{-11/3}\dot{f}_{\rm gw}\right)^{3/5}
\label{eq:Mchirp}
\end{equation}

In an X-ray binary system, however, the exchange of mass between the stellar components and mass lost from the system (as well as other effects giving rise 
to loss of $J_{\rm orb}$) affect the orbital period evolution and hence $\dot{f}_{\rm gw}$ cannot be evaluated using Equation~(\ref{eq:Jdotgw}).
In particular, for binaries where RLO results in a widening of the binary system, one has $\dot{f}<0$, which means that there is no real number solution to Equation~(\ref{eq:Mchirp}). Instead, I define the {\it dynamical} chirp mass, $\mathcal{M}_{\rm dyn}$:
\begin{equation}
\mathcal{M}_{\rm dyn} \equiv \left\{ \begin{array}{rcl}
     \displaystyle\frac{c^3}{G}\,\left(\frac{5}{96}\pi^{-8/3}\,(2f)^{-11/3}\;2\dot{f}\right)^{3/5}                      & {\rm for} & \dot{f} \ge 0 \\
    & & \\
     -\displaystyle\frac{c^3}{G}\,\left(\frac{5}{96}\pi^{-8/3}\,(2f)^{-11/3}\;|2\dot{f}|\right)^{3/5}                      & {\rm for} & \dot{f} < 0 \\
\end{array}\right.
\end{equation}
which will be negative for expanding orbits. That is, the orbital frequency, and hence the GW frequency, from expanding orbits will decrease and give rise to a negative chirp~\cite{kblk17}.

The reason for the change in the sign of orbital frequency (i.e. switching from a decreasing to an increasing orbital period) and the shape of computed UCXBs tracks (Figure~2) can be understood from the ongoing competition between GW radiation and orbital expansion caused by mass transfer/loss~\cite{stli17}. 
The peak at $f_{\rm gw}\simeq 5.5\;{\rm mHz}$ (corresponding to the minimum orbital period of $P_{\rm orb}\simeq 6.1\;{\rm min}$) coincides with the maximum value of the mass-transfer rate, $|\dot{M}_2|=10^{-6.8}\;M_{\odot}\,{\rm yr}^{-1}$.  
As the onset of RLO in the UCXB phase leads to a very high mass-transfer rate (Figure~1), an outward acceleration of the orbital size results from the small mass ratio between the two stars ($q\sim 0.1$), such that at some point the rate of orbital expansion dominates over that of orbital shrinking due to GW radiation. 

An analogy to the numerical computations of the described UCXB model can be made to RLO in double WD systems (see Figures S2--S5 in the Supplemental Material~\cite{SM}), i.e. the so-called AM~CVn binaries~\cite{hb00,kbs12} which constitute a main population of LISA sources~\cite{nyp01}.

\begin{figure*}
  \vspace{-1.2cm}\includegraphics[width=14cm]{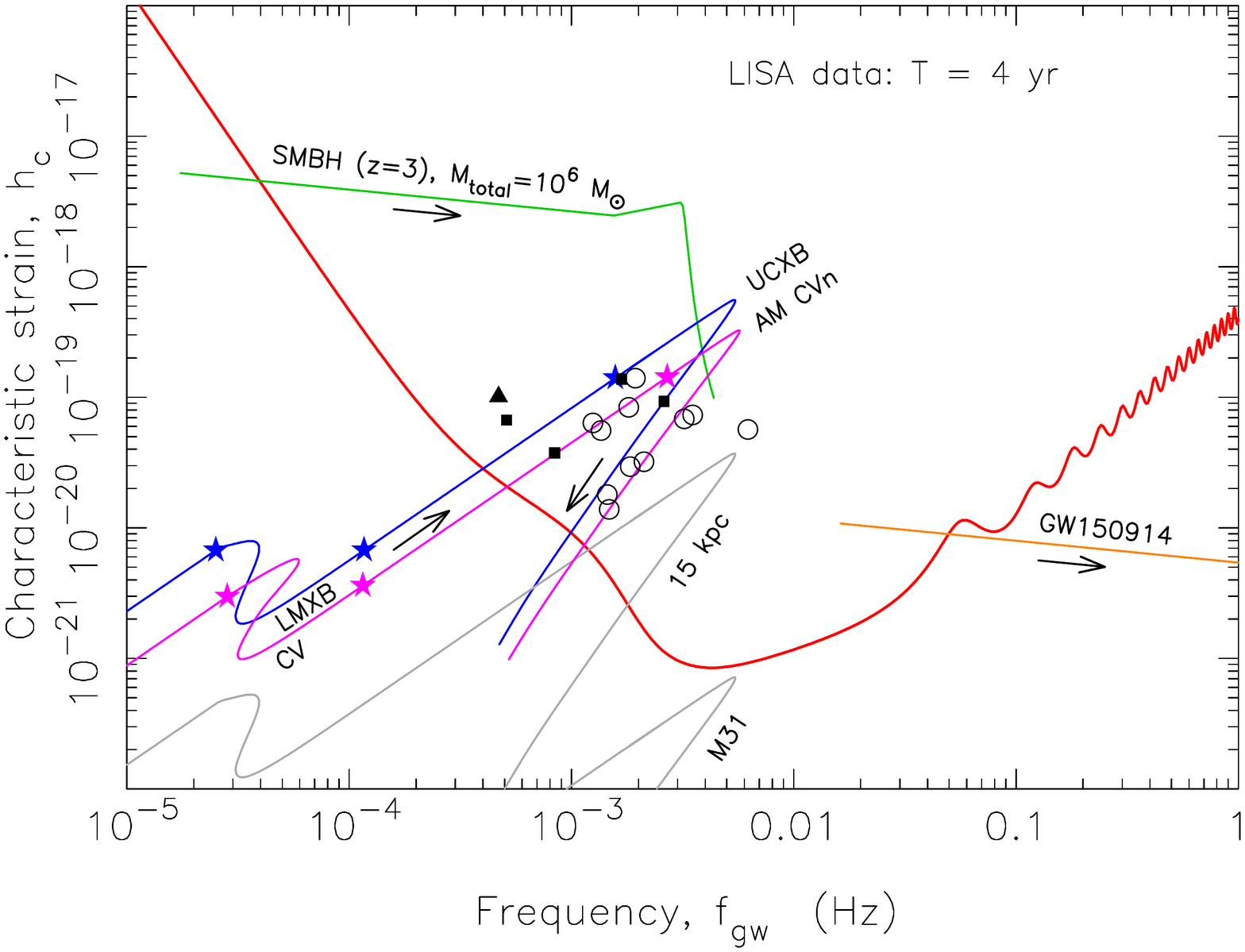}
  \caption{\label{fig3} Characteristic strain amplitude vs GW frequency for LISA. 
Evolutionary tracks for the UCXB system (blue) shown in Figures~1 and~2, and the example AM~CVn2 system (magenta, see Supplemental Material~\cite{SM}), at a distance of $d_L=1\;{\rm kpc}$. The stars represent (with increasing $f_{\rm gw}$): onset LMXB/CV stage, termination LMXB/CV stage, and onset UCXB/AM~CVn stage. The LISA sensitivity curve~\cite{cr18} (red line) is based on 4~yr of observations. 
The grey curves are for the UCXB at $d_L=15\;{\rm kpc}$ and $d_L=780\;{\rm kpc}$ (M31), respectively. Comparison tracks are shown for a super-massive black hole (SMBH) merger (green) and GW150914 (orange). Data from LISA verification sources~\cite{kks+18} include: detached double WD binaries (solid squares), AM~CVn systems (open circles) and a hot subdwarf binary (solid triangle).}
\end{figure*}

\paragraph{LISA observations.}
Figure~3 shows the characteristic GW strain amplitude calculated from the above evolutionary tracks for sources located at different distances with respect to the Solar System. The LISA sensitivity curve~\cite{cr18} based on 4~yr of observations is plotted for comparison (see Supplemental Material~\cite{SM}). The resulting signal-to-noise ratio (SNR) is above 100 out to distances of about 1~kpc, and for such a source located in the Andromeda galaxy (M31) at a distance of 780~kpc the peak characteristic strain is almost detectable. A comparison track calculated with MESA for a double WD system (AM~CVn) with component masses of $0.160\;M_{\odot}$ and $0.706\;M_{\odot}$ is included (see Supplemental Material~\cite{SM}), as well as computed tracks for a super-massive BH (SMBH) merger with a total mass of $10^6\;M_{\odot}$ at a redshift of $z=3$ (green line) and the last 4 years of in-spiral for the first LIGO event~\cite{aaa+16}, the double BH binary GW150914 (orange line), as it would appear in the LISA frequency band.

Derivations of individual component masses are not possible from LISA measurements alone, since higher order relativistic terms~\cite{hug09} to the quadrupole formula are needed (but not measurable) to break the degeneracy in component masses obtained from the observed chirp mass. To first post-Newtonian order (see section~5.2.2 in \citet{hug09}), the correction to the measured binary phase scales with $(v/c)^2$, where $(v/c)$ is the ratio between the relative orbital velocity and the speed of light. As an example, consider the UCXB model shown in Figure~2. Near the onset of the mass transfer from the WD to the NS (at which point the subsequent orbital evolution at latest deviates from pure GW radiation, here neglecting tidal effects) the GW frequency is $f_{\rm gw}$\,=\,1.63~mHz and $v\simeq$\,1068\;km\,s$^{-1}$, and thus $(v/c)^2\simeq 1.3\times 10^{-5}$. Such a small deviation in phase is not measurable with LISA.

\paragraph{A new method to determine NS masses.}
A tight correlation, however, exists between the orbital period and the mass of a helium WD which is produced in an LMXB system~\cite{sav87,ts99}. This correlation has been confirmed both observationally~\cite{vbjj05,tv14} and using the latest detailed binary stellar models including diffusion processes and rotational mixing~\cite{imt+16}. 
Since only post-LMXB NS\,+\,WD binaries with orbital periods less than about 9~hr are able to coalesce within a Hubble time (and thereby becoming visible LISA sources), the masses of all these WDs turn out to be the same within a narrow range ($M_{\rm WD}=0.162\pm 0.005\;M_{\odot}$, see Supplemental Material~\cite{SM}). 
This fortunate circumstance enables an accurate determination of NS masses within $\sim$~4\,\% (Figure~4), provided precise measurements of chirp masses in pre-UCXB systems.
For the best cases, it is found that the uncertainty of the measured $\mathcal{M}$ will be 0.5--1\,\% (see Supplemental Material~\cite{SM}). The resulting precise NS mass determinations may then yield a new upper mass limit of a NS accretor~\cite{afw+13} which helps to constrain the long-sought-after equation-of-state of NS matter~\cite{of16}.
A similar approach can be applied to infer the mass of the first-formed WD in a double WD system, which originates from stable RLO in a CV system.

\begin{figure}
  \vspace{-0.7cm}\includegraphics[width=10.5cm]{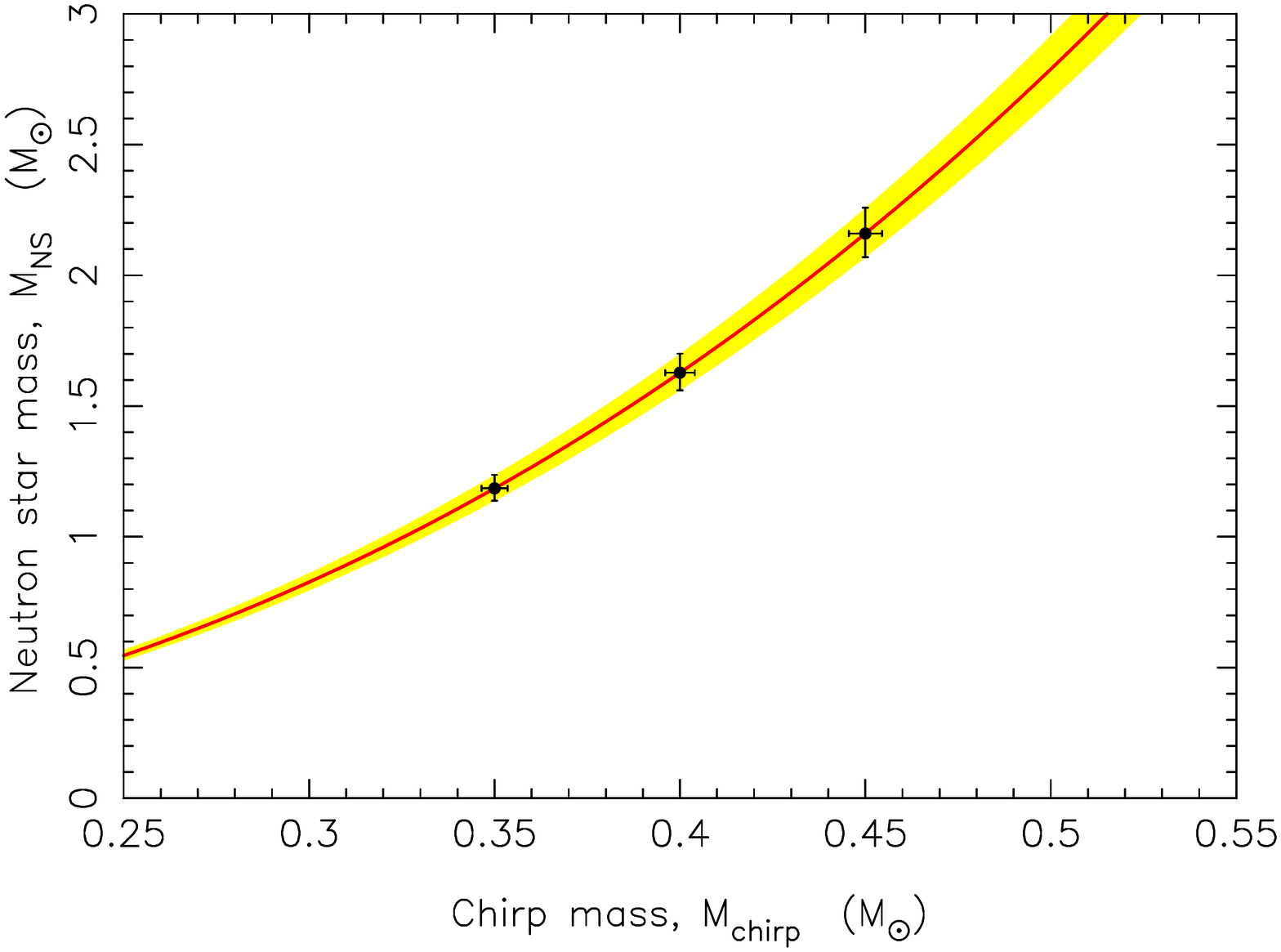}
  \caption{\label{fig4} NS mass vs measured chirp mass. 
For NS\,+\,WD binaries there is a unique correlation between orbital period and helium WD mass~\cite{sav87,ts99} after the LMXB stage and all such LISA (progenitor) systems have a well-defined WD mass of $0.162\pm 0.005\,M_\odot$ before the two compact objects coalesce and initiate mass transfer. Therefore, given a precisely measured chirp mass in such a binary (three examples indicated with an uncertainty of 1\,\%, see Supplemental Material~\cite{SM}) it is possible to derive the NS mass to an accuracy of about 4\,\%. The plotted relation is also valid for double WD systems (pre-AM~CVn systems) which evolved from a stable CV system~\cite{jrl87}.}
\end{figure}

A caveat is that LISA will only be able to measure $\dot{f}_{\rm GW}$ for nearby GW binaries with a very large SNR and which are close to their minimum orbital period where the rate of change in frequency is largest (see Figure~S1 in the Supplemental Material~\cite{SM}). 
However, this is where multi-messenger astronomy~\cite{llnc13} combining GWs and electromagnetic radiation is beneficial (Figure~1), including distance measurements of nearby sources using {\it GAIA} which can be combined with GW strain amplitude measurements to constrain $\dot{f}_{\rm gw}$~\cite{bkb+18}. 
Optical observations of the WD~\cite{hkb+12b} and searches for radio pulses from the NS can identify the nature of the LISA sources and help inferring the chirp mass by measuring $\dot{f}$. Similar targeted searches for radio pulsations from NSs in {\it Fermi} detected $\gamma$-ray sources have proven quite successful~\cite{aaa+09}.
Combined radio and optical observations of binary pulsars and WDs in close-orbit LISA (progenitor) sources, will enable further tests of e.g. WD formation, tidal effects and general relativity~\cite{afw+13}.

It is anticipated that measuring $\dot{f}_{\rm gw}$ is possible in about 25\,\% of the thousands of resolved LISA sources expected to be found~\cite{aab+12}. The precision of the measured values of $f_{\rm gw}$ and $\dot{f}_{\rm gw}$ increases over time, and a few sources may even have high enough SNR that allow for a measurement of $\ddot{f}_{\rm gw}$~\cite{aab+12}. Although $\dot{f}_{\rm gw}\simeq 0$ for sources {\it very} near to their minimum orbital period (which prevents a measurement of $\mathcal{M}_{\rm dyn}$), this epoch is short lasting (Figures~2b and S1) and will thus only affect a few systems.  
For UCXBs and AM~CVns, secular effects from tidal and mass-transfer interactions may introduce short-term variations in the measured values, but these effects will most likely not prevent detection of $\dot{f}_{\rm gw}$~\cite{sn09,kblk17} and thereby $\mathcal{M}_{\rm dyn}$. 

A remaining issue for refining the solutions presented here is calculating the exact evolution of $\dot{f}_{\rm gw}$ related to the torque balance arising from angular momentum advected from the donor to the accretion disk along with the transferred matter and the return of angular momentum from the disk to the orbit by means of a tidal torque between the outer disk and the donor~\cite{lp79,mns04,vnv+12}. 
Whereas detailed modelling of the accretion is disk is left for future studies, I performed trial computations with MESA including tidal effects, diffusion and rotational mixing of the WD, following~\citet{imt+16}.
The resulting change in entropy (inflated WD envelope) is found to be very limited (at the level of a few percent). Furthermore, the cooling properties of the WD also change when including diffusion and rotational mixing~\cite{imt+16} and at the onset of the UCXB stage $f_{\rm gw}$ is slightly {\it larger} (2.06~mHz vs 1.63~mHz). 

\paragraph{Number of Galactic NS\,+\,WD LISA sources.}
The number of UCXBs (and detached NS\,+\,WD systems prior to the UCXB phase) that LISA will detect is expected to be significantly smaller than the number of AM~CVn and detached double WD systems~\cite{nyp01}. Simple estimates based on known numbers of binary radio MSPs (see Supplemental Material~\cite{SM}) reveal nevertheless an expected LISA population of at least a hundred sources with NSs in the Milky Way. 

\paragraph{Dual-line gravitational wave system.}
With capabilities to calculate through two phases of mass transfer, the LMXB and the UCXB phases, it is possible to develop better models to follow the evolution of the accreting NSs and make improved theoretical predictions for their distribution of spin rates --- with applications to potential LIGO/Virgo detections of continuous high-frequency GWs from rapidly spinning NSs~\cite{hpp+15,aaa+18}. With a bit of luck, 
a Galactic dual-line GW frequency system can be detected from a combination of NS spin and orbital motion. The LIGO/Virgo detectors may detect a high-frequency GW signal ($f_{\rm gw}=2f_{\rm spin}$) from a rapidly spinning NS (note, recycled MSPs reside in these binaries) with some ellipticity, $\varepsilon$ and a resulting strain amplitude~\cite{hpp+15,aaa+18}:
\begin{equation}
  h_{\rm spin} = \frac{(4\pi)^2GI_{zz}f_{\rm spin}^2\;\varepsilon}{c^4\,d_L}
\end{equation}
where $I_{zz}$ is the principal moment of inertia and $d_L$ is the luminosity distance.
LISA may then measure the low-frequency GW signal ($f_{\rm gw}=2f_{\rm orb}$) arising from the orbital motion, $h_{\rm orb}$ with a strain amplitude given by~\cite{eis87}: 
\begin{equation}
  h_{\rm orb} = \left(\frac{32}{5}\right)^{1/2}\,\frac{(2\pi)^{2/3}G^{5/3}f_{\rm orb}^{2/3}\,\mathcal{M}^{5/3}}{c^4\,d_L}
\label{eq:h_orb}
\end{equation}
generated by a binary for an average orbital orientation and polarization. 
Combining these two expressions yields:
\begin{equation} 
  I_{zz}\,\varepsilon = \left(\frac{32}{80}\right)^{1/2}(2\pi)^{-4/3}\,G^{2/3}\left(\frac{f_{\rm orb}^{1/3}}{f_{\rm spin}}\right)^2\,\mathcal{M}^{5/3}\,\left(\frac{h_{\rm spin}}{h_{\rm orb}}\right)
\end{equation}
Once the right-hand-side of this equation is determined observationally, and assuming that the NS mass, $M_{\rm NS}$ can be determined from the chirp mass, $\mathcal{M}$ (under the assumption of $M_{\rm WD}$\,=\,0.162$\pm$0.005\,M\,$_\odot$), constraints can be made on the NS moment of inertia, and thus the NS  radius~\cite{rp94} (although only in combination with the ellipticity, $\varepsilon$) and thereby help pinning down the long-sought-after equation-of-state of NS matter.

\vspace{0.5cm}
\acknowledgments
The author is grateful for comments on the manuscript and many discussions with Norbert Wex. The author would also like to thank Norbert Langer, Alina Istrate, Rahul Sengar, Ed van den Heuvel, Laura Spitler and Lijing Shao for discussions on the formation and evolution of WDs and UCXBs, MESA, LISA sensitivity curves, and general comments. 

\vspace{0.2cm}
\paragraph{Supplemental Material.}
Supplemental material is included in the online version of this paper.
\nocite{pms+15,itl14,tv06,bdh12,imt+16,avk+12,npvy01,nyp01,kblk17,dbn05,dtwc07,btdn06,kbs12,rs85,mlp16,sa18,itla14,vbjj05,afw+13,hmg+13,kmw+14,nypv01,tcmr14,kks+18,ts02,sn14,mag08,cr18,ft00,eis87,aaa+16,abc+07,mhth05,lk12,kcb+18,hie+13,bkb+18}

\bibliographystyle{apsrev4-1}
\bibliography{tauris_refs}

\newpage
\newpage
\pagenumbering{roman}
\setcounter{page}{1}
\clearpage
\newpage


\section*{Supplementary material}

# SUPPLEMENTAL MATERIAL

## Disentangling Coalescing Neutron Star–White Dwarf Binaries for *LISA*

Thomas M. Tauris[1, 2, 3, *]

[1]*Argelander Institut für Astronomie, Auf dem Hügel 71, D-53121 Bonn, Germany*
[2]*Max-Planck-Institut für Radioastronomie, Auf dem Hügel 69, D-53121 Bonn, Germany*
[3]*Department of Physics and Astronomy, Aarhus University, Ny Munkegade 120, 8000 Aarhus C, Denmark*
(Dated: September 8, 2018)

## BINARY STELLAR EVOLUTION MODELLING

Binary stellar evolution models are calculated using the MESA code [11] (Modules for Experiments in Stellar Astrophysics, version 10108); from the initial MS + NS stage, through the two phases of mass transfer (the LMXB and the UCXB phases), until the computations cease (see below) with a planet orbiting an MSP (the recycled NS). The initial configuration used is a $1.40\,M_\odot$ donor star orbiting a $1.30\,M_\odot$ NS with an orbital period of 3.0 days. The MESA inlists are available upon request to the author, and the input physics is briefly described in the following. The NS is treated as a point mass. The chemical composition of the donor star is X = 0.6933, Y = 0.2863, Z = 0.0204. The mixing length parameter is set to $\alpha_{mix} = 2.0$ and convection is treated according to the Schwarzschild criterion. Orbital angular momentum changes due to magnetic braking and additional spin-orbit couplings [11] are included, as well as mass transfer/loss and GW radiation. The magnetic braking mechanism is evolved with an (ad hoc) index $\gamma = 4$. Using another strength for the magnetic braking would require a simple shift in the initial orbital period to reproduce the same system [42]. For the mass loss from the system, the isotropic re-emission model [43] is applied with $\alpha = 0$, $\beta = 0.70$, $\delta = 0$. That is, during RLO any wind-mass loss ($\alpha = 0$) from the donor star or production of a circumbinary disk ($\delta = 0$) is neglected. In addition, irradiation effects of the donor star via pulsar winds or photons are also ignored here (the latter effects have little effect on the overall UCXB evolution [44]). The effects of tides, diffusion and rotational mixing [26] were included in a couple of trial computations but were found to have a rather limited effect on the calculations. These effects are therefore not included in the default models presented here (although see Figure S5).

For the fraction of transferred material re-ejected from the vicinity of the NS accretor (via a jet, disk instabilities or a disk wind), $\beta = 0.70$ is chosen since MSP masses are often found to be relatively low, indicating inefficient accretion [45]. Hence the accretion efficiency of the NS is at most $\epsilon = 1 - \alpha - \beta - \delta = 0.30$, but limited to the Eddington mass-accretion rate of $1.5\times10^{-8}\ M_\odot\,\mathrm{yr}^{-1}$. Released gravitational binding energy of baryonic material accreted onto the NS is implemented, corresponding to a mass loss of 10 % in gravitational mass. The initial orbital period of the binary needs to be finetuned [42] such that the outcome is a detached NS + WD binary after magnetic braking and the LMXB phase.

The progenitor system [43] of the NS + MS binary (which is the starting point for the computations presented here) is a massive star ($10–25\,M_\odot$) orbited by a low-mass star (only slightly more massive than the Sun). Since the nuclear evolution timescale of the massive star, until it explodes in a supernova and produces the NS, is less than at most a few 10 Myr, it can be ignored here for determining the age of the UCXB system because of the Gyr timescale on which its low-mass MS star companion evolves.

A clear advantage of using a numerical stellar evolution code for the UCXB (and the AM CVn phase) is that it computes the structure of the WD donor self-consistently rather than relying on approximate expressions based on zero-temperature mass-radius relations [46, 3, 8]. Some previous studies have considered finite-temperature (specific entropy) effects or deviations from a simple fully convective structure and adiabatic evolution. However, these investigations approximated the evolution by using exposed cores of red-giant stars evolving from a second common-envelope phase [47] or applying single helium-star tracks [48]. Here, the helium WD donors are calculated via mass transfer in a progenitor binary which is then evolved coherently and followed through the subsequent UCXB (or AM CVn) stage. Hence, also the amount of hydrogen in the residual envelope after LMXB (or CV) detachment, and the rate at which it burns, is probed in detail with a code like MESA [26]. Including effects of (stable and unstable) nuclear burning on the surface of an accreting WD [49, 7] can be included in future work on AM CVn systems, but is not expected to change the longterm global evolution presented here.

The computation shown in Figures 1 to 3 terminates after ~14 Gyr because of issues with the equation-of-state of the ever-decreasing low-mass donor. The final system is composed of a $5.75\times10^{-3}\,M_\odot$ planet-like donor object, orbiting a $1.67\,M_\odot$ NS with $P = 1.17$ hr. It has been proposed that such systems may anyway become tidally disrupted [50] near this donor mass limit (which may then provide a promising channel to produce isolated radio MSPs with planets [51]).

## MASSES OF LOW-MASS HELIUM WDS

The masses of the helium WDs produced in LMXB systems which detach and later form UCXBs, when the orbit shrinks due to emission of GWs and the WD fills its Roche lobe and

TABLE I. Masses of helium WDs from detached LMXBs which produce UCXBs within a Hubble time, as modelled by MESA. The default system has: $M_{\text{donor}} = 1.4\,\text{M}_\odot$, $M_{\text{NS}} = 1.3\,\text{M}_\odot$, $Z = 0.0204$ and $\gamma = 4$. The effects on $M_{\text{WD}}$ by varying each of these parameters one by one are shown in columns 3 to 6. The last three columns are for the CV systems producing the AM CVn1 and AM CVn2 example models (see further below) and an UCXB model including tides, diffusion and rotational mixing [26]. The bottom row states the initial value of $P$ for each model.

| | default | $M_{\text{donor}} = 1.1\,\text{M}_\odot$ | $M_{\text{NS}} = 1.7\,\text{M}_\odot$ | $Z = 0.005$ | $\gamma = 5$ | CV1 | CV2 | UCXB$_{\text{Istrate}}$ |
|---|---|---|---|---|---|---|---|---|
| $M_{\text{WD}}$ (M$_\odot$) | 0.162 | 0.157 | 0.163 | 0.166 | 0.158 | 0.162 | 0.160 | 0.163 |
| $P_{\text{init}}$ (days) | 3.005 | 2.860 | 2.700 | 2.070 | 3.500 | 3.640 | 3.9482 | 3.02 |

initiates mass transfer to the NS, are all found to be between $0.157\,\text{M}_\odot$ and $0.166\,\text{M}_\odot$ (i.e. $0.162 \pm 0.005\,\text{M}_\odot$). This result is obtained by investigating the evolution of a number NS + MS systems using MESA, in which key input parameters have been altered (see Table 1): the masses of the MS star ($M_{\text{donor}}$) and the NS star ($M_{\text{NS}}$), the metallicity ($Z$) and the magnetic braking index parameter, $\gamma$. Only by changing the metallicity to an extremely low content ($Z = 0.001$) for Galactic sources, or applying a very small $\gamma$-value ($\gamma = 2$), larger values of $0.172\,\text{M}_\odot$ and $0.174\,\text{M}_\odot$, respectively, are found. However, in the latter case, the UCXB systems do not form within a Hubble time and these models can be disregarded. CV systems produce AM CVn binaries with identical helium WD donor star masses [52] (see further below).

Observations of low-mass helium WDs orbiting radio MSPs in tight orbits with $P < 9$ hr confirm that they have masses of roughly $0.16 \pm 0.02\,M_\odot$ (see [10] and references therein), although model-atmosphere fits to determine their temperature and surface gravity, combined with a mass-radius relation to estimate the WD mass, have some uncertainties involved [24, 27]. Similar masses are found for 5 out of 6 helium WD companions to binary MSPs found in globular clusters with circular orbits (eccentricities $< 10^{-4}$) and $P < 15$ hr [10]. Further evidence for masses of $0.16 \pm 0.02\,M_\odot$ is found among the pulsating low-mass helium WDs [53] and, for example, the detached double WD binary NLTT 11748 [54]. It should be cautioned, however, that not all close-orbit WD + WD (or double WD) systems may have formed via stable mass-transfer in an LMXB (or CV) system. In fact, the majority of the LISA sources which are double WD systems are likely to host two relatively more massive WDs which formed via double common-envelope evolution [55, 56].

For the calculated CV system plotted in Figure 3 (which has a WD accretor and leads to the AM CVn2 sample system (see below), the initial binary consists of a WD + MS binary with masses of $M_{\text{WD}} = 0.50\,\text{M}_\odot$ and $M_2 = 1.4\,\text{M}_\odot$, respectively, and an initial orbital period of ~3.95 days. For the WD accretor, an Eddington accretion limit of $2.6 \times 10^{-7}\,\text{M}_\odot\,\text{yr}^{-1}$ is assumed and $\beta = 0.80$. As an AM CVn system with a total mass of less than $0.90\,\text{M}_\odot$ this system resembles a number of LISA verification binaries [20]. The final outcome of this calculation is a $6.8 \times 10^{-3}\,\text{M}_\odot$ planet-like donor object, orbiting a $0.74\,\text{M}_\odot$ WD with an orbital period of 1.06 hr.

## CHIRP MEASUREMENT

LISA will only be able to measure $\dot{f}$ for nearby GW binaries with a very large signal-to-noise ratio (SNR) and which are close to their minimum orbital period where the rate of change in frequency is largest. The resolution can be estimated by considering the change in orbital phase, $\phi(t)$ (orbital cycles) caused by $\dot{f}$ and a simple Taylor expansion:

$$\phi(t) = \phi_0 + f(t - t_0) + \frac{1}{2}\dot{f}(t - t_0)^2 + \dots \quad (1)$$

where $T = t - t_0$ is the total observation time. For a given SNR, assume that phase shifts of the order of 1/SNR orbits are measurable. This yields a limit on the minimum value of $\dot{f}_{\text{gw}}$ (recalling $\dot{f}_{\text{gw}} = 2\dot{f}$):

$$\dot{f}_{\text{gw,min}} \sim \frac{4}{T^2}\frac{1}{\text{SNR}} \simeq 2.5 \times 10^{-18}\left(\frac{100}{\text{SNR}}\right)\left(\frac{4\,\text{yr}}{T}\right)^2\,\text{Hz}\,\text{s}^{-1} \quad (2)$$

in agreement with more detailed estimates [57] and corresponding to detection of a minimum $|\dot{f}_{\text{gw}}| \simeq 2.5 \times 10^{-17}\,\text{Hz}\,\text{s}^{-1}$ for a 4 year LISA mission and a GW source with SNR = 10. By comparison to the calculated UCXB tracks presented here (Figure 2), this means that only nearby UCXBs close to their minimum orbital period (peak frequency) will have a measurable chirp signal. The amount of time binary GW sources spend near the peak frequency is relatively short and statistically most LISA sources will not have orbits that enable a dynamical chirp mass measurement. In any case, (at latest) after the onset of RLO, before reaching the peak frequency, the measured $\dot{f}$ does not arise from GW radiation alone. Mass transfer and tidal effects [8, 58] will also affect the measured value of $\dot{f}$ and thus $\mathcal{M}_{\text{dyn}} \neq \mathcal{M}$, even while $\dot{f} > 0$.

To complete the argument of how well NS masses can be determined, we consider the uncertainty of the measured chirp mass. From Equation (3) in the main paper, one derives:

$$\frac{\Delta\mathcal{M}}{\mathcal{M}} = \frac{11}{5}\frac{\Delta f_{\text{gw}}}{f_{\text{gw}}} + \frac{3}{5}\frac{\Delta\dot{f}_{\text{gw}}}{\dot{f}_{\text{gw}}} \quad (3)$$

This expression can be combined with error estimates of $\Delta f_{\text{gw}}$ and $\Delta\dot{f}_{\text{gw}}$. Using equations (10) and (11) in [57] yields:

$$\frac{\Delta\mathcal{M}}{\mathcal{M}} \simeq 3.8 \times 10^{-7}\left(\frac{100}{\text{SNR}}\right)\left(\frac{4\,\text{yr}}{T}\right)\left(\frac{1\,\text{mHz}}{f_{\text{gw}}}\right)$$
$$+ 1.6 \times 10^{-2}\left(\frac{100}{\text{SNR}}\right)\left(\frac{4\,\text{yr}}{T}\right)^2\left(\frac{10^{-16}\,\text{Hz}\,\text{s}^{-1}}{\dot{f}_{\text{gw}}}\right) \quad (4)$$

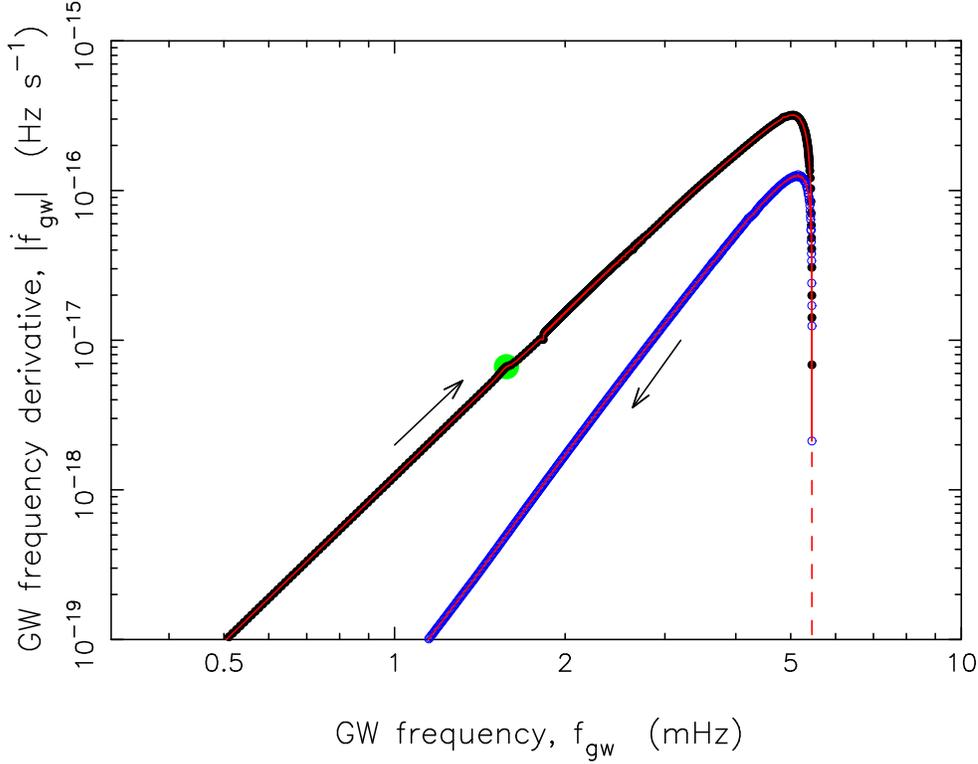

**Fig. S1.** GW frequency derivative, $|\dot{f}_{gw}|$ as a function of GW frequency, $f_{gw}$ for the UCXB shown in Figures 1 and 2. The black colored points correspond to the inbound leg (orbital shrinking) and the blue points correspond to the outbound leg (orbital expansion, after reaching the orbital period minimum). Each point represents a binary stellar MESA model. The green solid circle indicates the onset of the UCXB stage.

and where the SNR is inversely proportional to the distance. As an optimal example, consider a system with: SNR = 100, $T = 4$ yr, $f_{gw} = 5$ mHz and $\dot{f}_{gw} = 3.2 \times 10^{-16}$ Hz s$^{-1}$, which yields $\Delta\mathcal{M}/\mathcal{M} \simeq 0.005$, i.e. an uncertainty of only 0.5 %. For most LISA sources, the error on $\Delta\mathcal{M}$ is completely dominated by the second term in Equation (4).

Figure S1 shows the evolution of $|\dot{f}_{gw}|$ as a function of $f_{gw}$, for the UCXB calculation plotted in Figures 1 and 2 in the main paper. On the inbound leg of the orbital evolution (black points) the value of $\dot{f}_{gw}$ is positive, whereas it is negative (giving rise to a negative dynamical chirp mass) on the outbound leg (blue points). The maximum value of $\dot{f}_{gw}$ is close to the maximum value of $f_{gw}$, but not exactly given that $\dot{f}_{gw}$ changes sign at the orbital period minimum. For a given value of $f_{gw}$, the value of $|\dot{f}_{gw}|$ is larger on the inbound leg compared to the outward leg.

### NUMERICAL NOISE IN COMPUTATIONS

In producing Figures 2b and S1, two data points out of 1547 were removed due to numerical noise in the computations. Furthermore, a noise filter was applied to smoothen any discontinuous changes in the computed chirp mass exceeding 2% between two consecutive data points of the model. The plotted discontinuity in the derived value of $\mathcal{M}_{dyn}$ towards the termination of the LMXB phase (dashed line in Figure 2b) is caused by switching off magnetic braking at this epoch. This causes a discontinuity in $\dot{J}_{orb}$ and therefore in $\dot{f}_{gw}$.

### LISA SENSITIVITY CURVE AND STRAIN AMPLITUDES

In the following paragraph, the GW frequency is denoted by $f$. The strain data stream observed at the detector, $s(t)$ is a combination of a true GW strain, $h(t)$ and noise, $n(t)$, i.e.: $s(t) = h(t) + n(t)$. The SNR is given by [17]:

$$\rho^2 = \left(\frac{S}{N}\right)^2 = 4 \int_0^\infty \frac{|\tilde{h}(f)|^2}{S_n(f)}\,df = 4 \int_{f=0}^\infty \frac{f|\tilde{h}(f)|^2}{S_n(f)}\,d(\ln f) \tag{5}$$

where $\tilde{h}(f)$ is the Fourier transform of the signal waveform given by:

$$\tilde{h}(f) \equiv \int_{-\infty}^\infty h(t)\,e^{-i\omega t}\,dt \tag{6}$$

and the noise power spectral density for a frequency resolution $\Delta f$ (roughly equal to $1/T$) is given by:

$$S_n(f) = 2\,\langle|\tilde{n}(f)|^2\rangle\,\Delta f \tag{7}$$

In the LISA community, the above $S_n(f)$ is called $P_n(f)$, and hence the full LISA sensitivity curve (strain spectral density) is given by [19]:

$$S_n(f) = \frac{P_n(f)}{\mathcal{R}(f)} + S_c(f) \tag{8}$$

where $\mathcal{R}(f)$ is the signal response (transfer) function which relates the power spectral density of the incident GW signal to the power spectral density of the signal recorded in the detector. Hence, $\mathcal{R}(f)$ is a function of detector geometry and antenna patterns. The last term in the above equation, $S_c(f)$, is the Galactic confusion noise from unresolved binaries [3]. The resulting LISA sensitivity curve plotted as a red line in Figure 3 is taken from Cornish and Robson [19].

For the computed tracks of the UCXB and AM CVn sources plotted in Figure 3, the change in frequency is small during the lifetime of the LISA mission. The maximum rate of change is $\dot{f}_{gw}^{max} = 3.2 \times 10^{-16}$ Hz s$^{-1}$ which corresponds to $\Delta f_{gw} = 4.0 \times 10^{-8}$ Hz in 4 yr. Thus, one can consider these sources as being monochromatic. For such sources with a well-defined $h(t)$, one can calculate the characteristic strain, $h_c$ [59] by accumulating the power in the signal over many cycles, $N_{cycles}$ for an observation time, $T$:

$$h_c \approx \sqrt{N_{cycles}} \ \sqrt{2} \, h_0 = \sqrt{2 f_{gw} \, T} \ h_0 \tag{9}$$

where

$$h_0 = \left(\frac{32}{5}\right)^{1/2} \ \frac{\pi^{2/3} G^{5/3} f_{gw}^{2/3} \mathcal{M}^{5/3}}{c^4 \, d_L} \tag{10}$$

is the GW amplitude [40] generated by a binary at luminosity distance, $d_L$ for an average orbital orientation and polarization. For the estimated GW strain amplitudes for the UCXBs in Figure 3, it is assumed that $T = 4$ yr. Along the inbound leg of these tracks the sources evolve with a slope of $h_c \propto f_{gw}^{7/6}$.

A couple of evolving sources are plotted in Figure 3 for comparison (i.e. binaries for which $f_{gw}$ changes significantly over the LISA observation time). Two examples are given for the last 4 yr of in-spiral for: i) a hypothetical super-massive black hole (SMBH) binary with a total mass of $10^6 \, M_\odot$ at redshift $z = 3$, and ii) GW150914 [1]. For these systems, after correcting for the redshift, the waveform model of the original phenomenological model called PhenomA [60] were implemented [19] for the full inspiral-merger-ringdown (SMBH) and the inspiral part (GW150914), respectively. These sources evolve with $h_c \propto f_{gw}^{-1/6}$ during the inspiral before the merger event.

## NUMBER OF GALACTIC UCXBS AND DETACHED CLOSE-ORBIT NS + WD SYSTEMS

The four tightest binaries known to contain a radio MSP and a low-mass helium WD companion are [61]: PSRs J0348+0432, 0751+1807, J1036−8317 and J1738+0333. All of these systems will coalesce within

a Hubble time and produce an UCXB (and a LISA source, slightly before and during the UCXB stage). Their distances are 2.10, 1.11, 0.93 and 1.47 kpc, respectively [61]. Assume a number density of such systems of about 1 kpc$^{-3}$ and a scale height above the Galactic disk of about 1 kpc (these systems have significantly larger scale heights than their progenitor stars due to the kinematic effect of the supernova explosion when the NS forms [43]). Furthermore, assume the Galactic disk being uniform with a radius of 10 kpc. This leads to a Galactic disk population of about 300 of such binary MSP systems. However, we must include the beaming effects [62], which for MSPs is about 30 per cent of the sky, leading to a total of 1000 Galactic binary MSPs. A detached NS + WD system which coalesces within a Hubble time has a typical age of 5 to 10 Gyr when it is formed, and remains detached for a few Gyr before the UCXB phase. However, it is only visible as a GW source in the LISA band for $f_{gw} > 0.4$ mHz (Figure 3), corresponding to the last 100 Myr before the UCXB phase (Figure 2). Thus, it is estimated that of order 50 detached NS + WD systems are observable with LISA. After the onset of the UCXB phase, a similar number (possibly larger by a factor of 2) of sources are detectable. Although these UCXBs remain relatively tight on a Gyr timescale, their dynamical chirp masses and GW amplitudes become smaller as the WD donor, and the system, loses mass (Figures 2 and 3).

The final numbers are likely to be somewhat larger than derived above since not all NS + WD systems host an active radio pulsar. Furthermore, a few systems located in globular clusters (some of which formed via a stable LMXB, thus obeying the orbital period–mass relation for helium WDs) will also contribute to the detected GW population [63]. For comparison, the number of observed Galactic UCXBs in the X-ray band is about 15, see ref. [14]. However, many selection effects are at work (such as the transient nature of many of these systems, the design of the X-ray surveys and the interstellar absorption from gas and dust), which means that a direct comparison to the above numbers is not trivial.

## WD + WD SYSTEMS AND AM CVN SYSTEMS

The Figures S2 to S5 illustrate the calculated MESA models of two AM CVn systems and their comparison to the UCXB model. The modelling of these two sample AM CVn systems is now described in more detail.

### AM CVn1

The progenitor system is a $1.40 \, M_\odot$ MS star orbiting a $0.7 \, M_\odot$ carbon-oxygen WD (treated as a point mass) with an initial orbital period of 3.64 days. After orbital decay caused by magnetic braking and the subsequent CV phase, the system detaches again with a $0.162 \, M_\odot$ helium WD orbiting a $1.07 \, M_\odot$ WD with an orbital period of 4.76 hr. Over the next 4.1 Gyr, this double WD system spirals in further due to

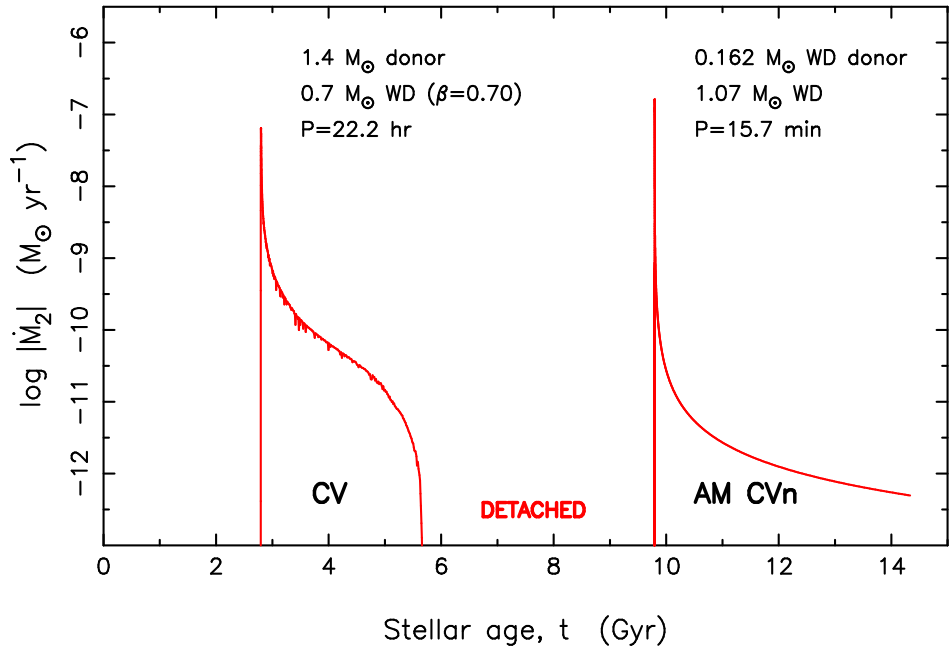

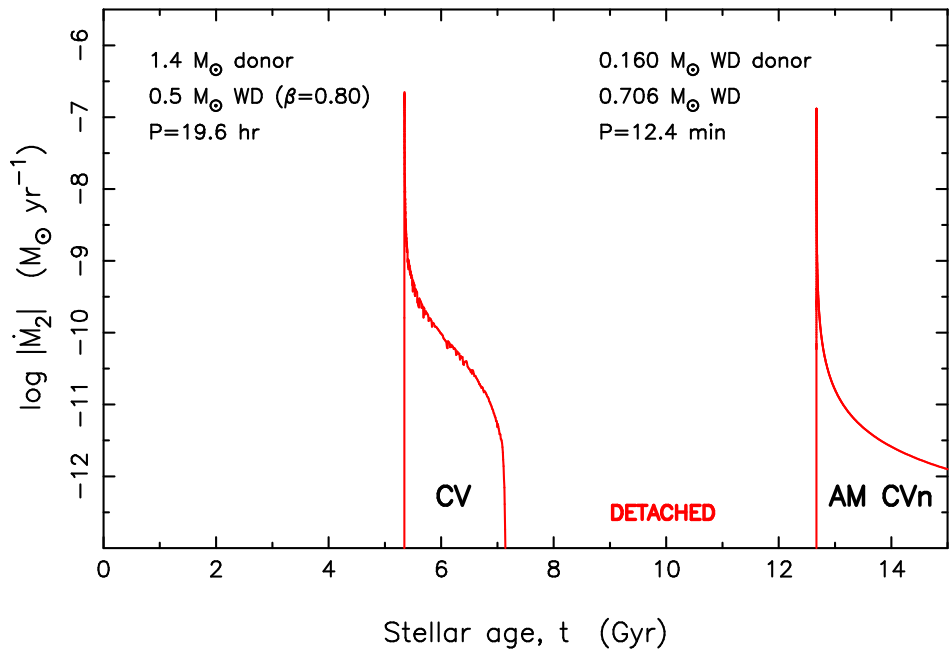

**Fig. S2.** Mass-transfer rate as a function of stellar age for two CV and AM CVn systems. The donor and the accretor star masses, as well as the orbital period, are stated at the onset of each mass-transfer phase. The top (bottom) panel shows the evolution leading to the AM CVn1 (AM CVn2) system, see text.

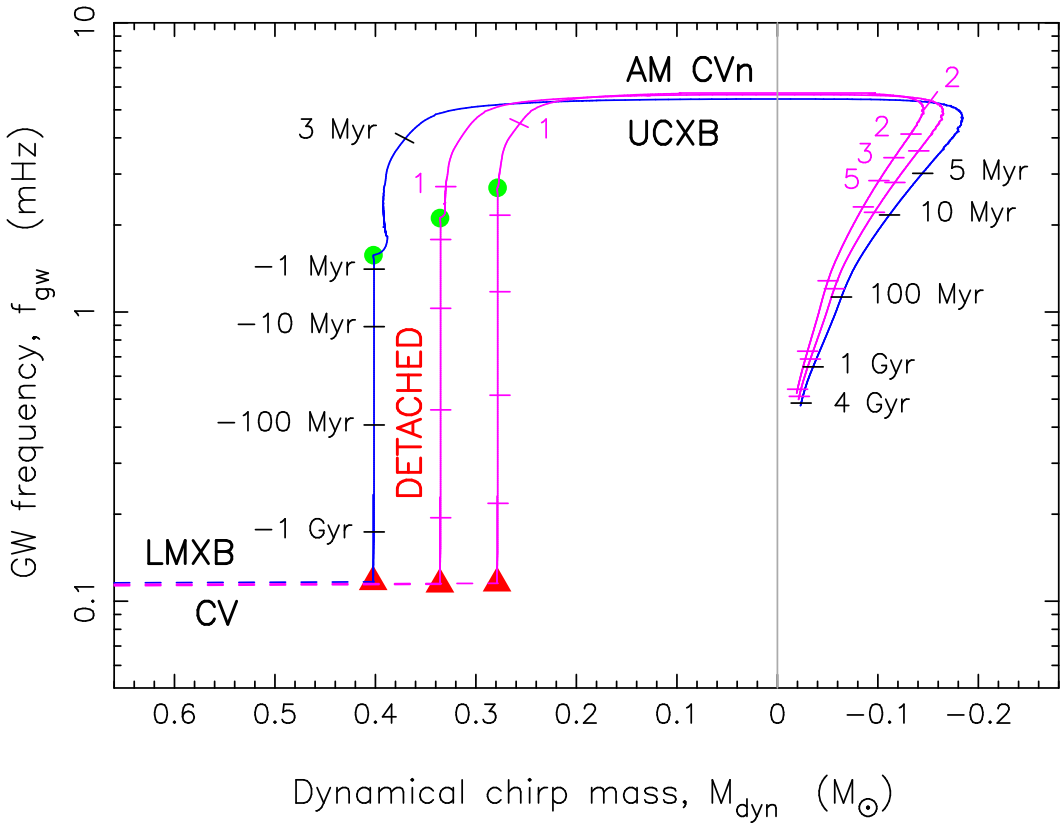

**Fig. S3.** GW frequency vs dynamical chirp mass for the UCXB and the two AM CVn systems. The blue track is for the UCXB system in Figures 1 and 2. The magenta tracks are the two AM CVn systems (Figure S2). The end points of the first mass-transfer phases (LMXB and CV) are indicated by red triangles; the starting points of the second mass-transfer phases (UCXB and AM CVn) are indicated by green circles. The time marks along the AM CVn tracks are for the same values as indicated for the UCXB system, unless stated otherwise (in Myr). Time zero is defined at the onset of the second mass-transfer phase.

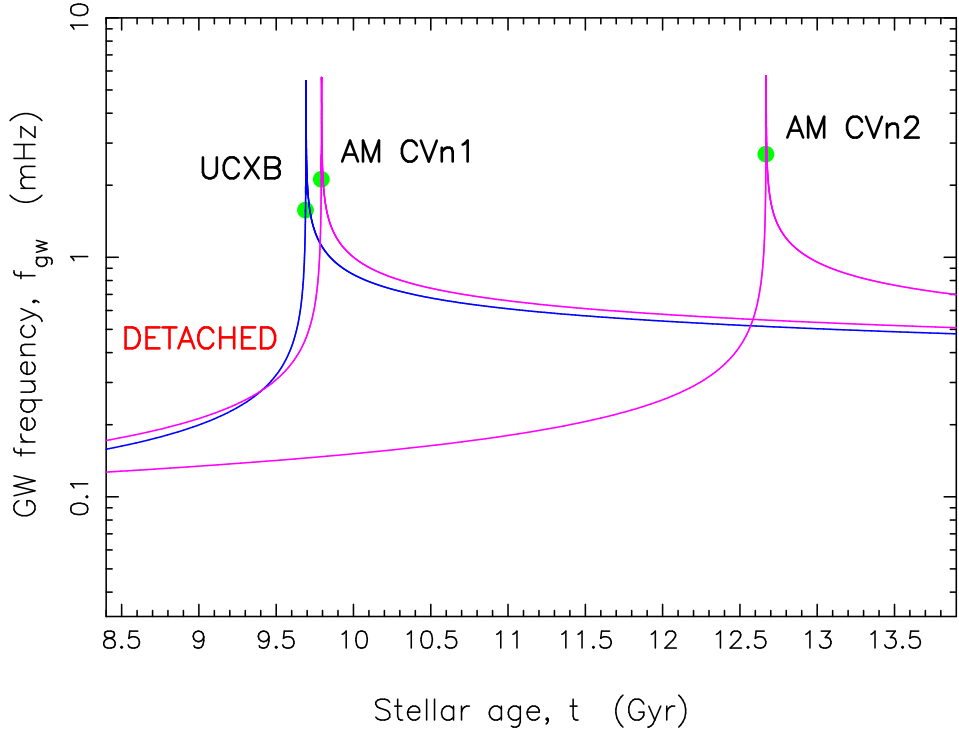

**Fig. S4.** GW frequency vs stellar age for the UCXB and the two AM CVn systems. The blue track is the UCXB system in Figure 2a in the main paper. The magenta tracks are the two AM CVn systems described in detail here in the Supplemental Material. The maximum GW frequencies in these three examples are 5.45 mHz (UCXB), 5.64 mHz (AM CVn1) and 5.72 mHz (AM CVn2), corresponding to minimum orbital periods of 6.12 min, 5.91 min and 5.83 min, respectively. The frequency at the onset of the RLO (green circles) depends on the temperature of the low-mass helium WD donor star ($T_{eff}$ = 10 850 K, 9 965 K and 8 999 K, respectively).

emission of low-frequency GWs with a constant chirp mass, $\mathcal{M} = 0.335 \, M_\odot$, until the low-mass helium WD fills its Roche lobe (at $P = 15.7$ min and $T_{eff} = 9\,965$ K) and initiates mass transfer to the carbon-oxygen WD and the binary becomes observable as an AM CVn system. The final calculated model is composed of a $6.26\times10^{-3} \, M_\odot$ planet-like donor object, orbiting a $1.11 \, M_\odot$ WD with an orbital period of 1.12 hr and a stellar age of 14.3 Gyr.

## AM CVn2

The aim of this model is to reproduce some of the known detached double WDs systems and AM CVn binaries which have lower WD masses [20] (i.e. total system masses below $0.9 \, M_\odot$). To achieve this, a progenitor system is chosen with a $1.40 \, M_\odot$ MS star orbiting a $0.5 \, M_\odot$ carbon-oxygen WD with an initial orbital period of 3.9482 days. Assuming rather inefficient accretion onto the WD ($\beta = 0.80$), the system detaches after the CV phase with a $0.160 \, M_\odot$ helium WD orbiting a $0.706 \, M_\odot$ carbon-oxygen WD with an orbital period of 4.81 hr. Over the next 5.5 Gyr, the detached system spirals in further due to GW radiation with a constant chirp mass,

$\mathcal{M} = 0.278 \, M_\odot$, until the low-mass helium WD fills its Roche lobe (at $P = 12.4$ min and $T_{eff} = 8\,999$ K) and initiates mass transfer to the carbon-oxygen WD as an AM CVn system. The final calculated model is composed of a $6.81\times10^{-3} \, M_\odot$ planet-like donor object, orbiting a $0.736 \, M_\odot$ WD with an orbital period of 1.06 hr and a stellar age of 17.4 Gyr.

Figure S2 shows the mass-transfer rate as a function of stellar age for the two CV systems evolving into the AM CVn systems (AM CVn1 and AM CVn2) described here in the Supplemental Material. The figure can be compared to Figure 1 in the main text for an LMXB system evolving into an UCXB system.

Figure S3, displaying GW frequency vs dynamical chirp mass, is similar to Figure 2b in the main paper, but includes also the two AM CVn models. The shapes of these three tracks are rather similar. However, a significant difference is seen in the dynamical chirp mass (between $0.401 \, M_\odot$ and $0.278 \, M_\odot$) for the detached systems which reflects the difference in mass between an accreting NS (UCXB) and an accreting WD (AM CVn).

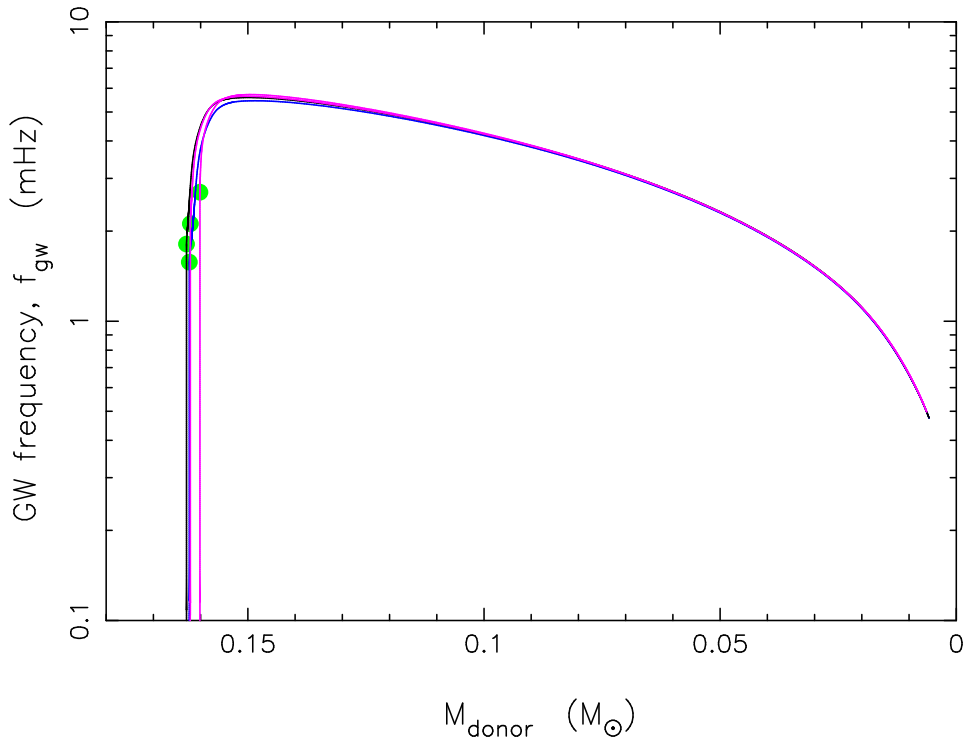

**Fig. S5.** GW frequency vs donor star mass for the UCXB and the two AM CVn systems shown in Figures S3 and S4. The evolution is from left to right. The green circles indicate the onset of RLO. The three tracks (same colors as in Figures S2 and S3) overlap very closely after RLO, and the plotted relation is clearly independent of the mass of the accreting compact object (here between 0.706 and 1.63 $M_\odot$ – see text). In addition, a fourth model is plotted (in black color) which is an UCXB model where tides, diffusion and rotational mixing are included [26].

In Figure S4 is plotted GW frequency vs stellar age (since the ZAMS) for both the UCXB and the two AM CVn models. The value of $f_{gw}$ at the onset of the UCXB or AM CVn stage is mainly determined by the temperature (and thus the age) of the helium WD donor which dictates its radius [10]. For this reason, zero-temperature mass-radius relations are not applicable to study these models. The systems spend about 11 Myr (UCXB), 14 Myr (AM CVn1) and 18 Myr (AM CVn2) emitting GWs with $f_{gw} > 2$ mHz where LISA is particularly sensitive, see Figure 3 in the main paper.

Figure S5 shows the evolution in GW frequency as a function of decreasing donor star mass for the UCXB and the two AM CVn models displayed in Figures S3 and S4. An additional UCXB model calculated including tides, diffusion and

rotational mixing [26] is shown as well (black color). The relation is remarkably independent of the mass of the accreting compact object, which for these three models varies between 0.706 $M_\odot$ (AM CVn2) and 1.63 $M_\odot$ (UCXB). Hence, by measuring $f_{gw}$ with LISA, one can infer the donor star mass [31] if, and only if, the observed GW source descends from stable RLO in a CV system or an LMXB.

---

\* tauris@phys.au.dk



\end{document}